# Phase-locked indistinguishable photons with synthesized waveforms from a solid-state source

Clemens Matthiesen[1], Martin Geller[1,2], Carsten H. H. Schulte[1], Claire Le Gall[1], Jack Hansom[1], Zhengyong Li[3], Maxime Hugues[4], Edmund Clarke[4] & Mete Atatüre[1,*]

[1]*Cavendish Laboratory, University of Cambridge, Cambridge CB3 0HE, United Kingdom*

[2]*Fakultät für Physik and CENIDE, Universität Duisburg-Essen, Duisburg 47048, Germany*

[3]*Key Laboratory of Luminescence and Optical Information of Ministry of Education, Beijing Jiaotong University, Beijing 100044, China*

[4]*EPSRC National Centre for III-V Technologies, University of Sheffield, Sheffield, S1 3JD, United Kingdom*

**Resonance fluorescence in the Heitler regime provides access to single photons with coherence well beyond the Fourier transform limit of the transition, and holds the promise to circumvent environment-induced dephasing common to all solid-state systems. Here we demonstrate that the coherently generated single photons from a single self-assembled InAs quantum dot display mutual coherence with the excitation laser on a timescale exceeding 3 seconds. Exploiting this degree of mutual coherence we synthesize near-arbitrary coherent photon waveforms by shaping the excitation laser field. In contrast to post-emission filtering, our technique avoids both photon loss and degradation of the single photon nature for all synthesized waveforms. By engineering pulsed waveforms of single photons, we further demonstrate that separate photons generated coherently by the same laser field are fundamentally indistinguishable, lending themselves to creation of distant entanglement through quantum interference.**

[*] ma424@cam.ac.uk.

Significant progress has been reported within quantum information science for quantum-dot (QD) spins as stationary qubits[1,2] including state preparation[3,4], long spin coherence times[5,6], ultrafast optical manipulation capabilities[7,8] and single-shot read-out[9]. A successful realization of a solid-state quantum network relies on scalable entangling gates between individual spins. Non-local entanglement based on photon quantum interference[10] has been identified as the most promising and robust approach[11]. The robustness stems from the probabilistic nature of entanglement creation, placing fewer constraints on detection efficiencies, but relies crucially and fundamentally on indistinguishable photons from separate spins. For QDs the quality of photons as flying qubits, however, has remained systematically below par due to detrimental effects of the solid-state environment on the photon generation process casting a major challenge on this roadmap today.

Resonance fluorescence, recently demonstrated for self-assembled InAs QDs[12,13,14,15], avoids optically exciting the host material and therefore suppresses these dynamics, however the ambient charge fluctuations, for example, can still broaden a transition via spectral wandering[16]. Deviation from a transform-limited photon wavepacket renders the use of QDs in a quantum optical network problematic, necessitating further progress in understanding these dynamics, as well as suppressing their effects. Elastic scattering from an optical transition[17] represents an alternative approach to high quality photon generation, which avoids significant population in dephasing-prone excited states. Unlike the spontaneously emitted photons, the elastically scattered photons in the Heitler regime are phase-locked to the excitation laser with ideally infinite mutual coherence, accentuating their potential for quantum interference[18]. In well-isolated experimental systems, such as trapped atoms and ions, this mutual coherence was observed via an optical heterodyning technique[19], which allowed studying the broadening of elastic scattering due to motional and vibrational dynamics[20, 21], when the species are driven by a single frequency laser. For semiconductor QDs elastic scattering in resonance fluorescence was observed recently[22, 23], ruling out strong fast dephasing of the optical transition. The effect of environment dynamics, apparent through spectral wandering for instance, can still limit the mutual coherence between the excitation laser and the photons generated in the Heitler regime.

Here, we demonstrate the suitability of coherently scattered QD photons as flying qubits in quantum optical networks. First, phase-sensitive heterodyning measurements are employed to determine the extent to which phase-locking to the excitation laser occurs. The sub-hertz mutual coherence indicates that spectral diffusion is not significant in our sample on the measurement timescale and allows us to synthesize the waveform of the coherently scattered single photons. The accessible bandwidth for waveform synthesis is restricted only by the mutual coherence for low frequencies, and the QD transition linewidth for high frequencies. Finally, we demonstrate through two-photon interference that the coherent QD photons are fundamentally indistinguishable, attesting their potential in quantum networks.

## Results

**Mutual coherence of QD photons and the excitation laser.** In the experiments reported here we excite resonantly and collect the resonance fluorescence from the neutral exciton transition of InAs/GaAs self-assembled QDs embedded in a Schottky heterostructure at 4.2 K (see Methods). The resonance fluorescence is isolated from the excitation laser with a signal to background ratio better than 100:1 using a cross-polarisation technique[14, 22]. The QD transition is Stark-tunable by 80 GHz around 315315 GHz (951 nm). Spectral wandering broadens the $2\pi \times 250$-MHz radiative linewidth noticeably only on timescales of seconds. Figure 1 (a) shows the experimental arrangement used to perform optical heterodyning measurements (see Methods). QD resonance fluorescence at frequency $\nu$ is superimposed with a strong local oscillator originating from the excitation laser but shifted in frequency by $\delta\nu \approx 210$ kHz. This frequency difference results in a beat note in the difference signal of two balanced photodiodes (PD1,2). A spectrum analyser calculates the power spectrum via a fast Fourier transform (FFT) algorithm. The linewidth and the lineshape of the power spectrum reflect directly the phase stability, i.e. the mutual coherence, between the two fields during the course of the measurement. In particular, periodic dynamics of the relative phase appear as sidebands and aperiodic fluctuations appear as broadening of the spectrum. Figure 1 (b) displays a representative raw spectrum with a strong peak at the beating frequency. The inset shows that a Gaussian lineshape with a full-width-at-half-maximum, $\Delta\nu_{FWHM}$, of 299±9 mHz fits the data of 5-second continuous acquisition. This raw spectrum is dominated by the system response resolution of 200 mHz for the measurement settings. A longer acquisition time yields higher spectral resolution, hence narrower linewidths (e.g. 172±12 mHz for 10-second acquisition time), but phase fluctuations induced by the mechanical instability of the setup reduce the signal quality and limit the obtainable linewidth (see Supplementary Note 1 and Supplementary Fig. S1). Nevertheless, even these raw linewidths indicate that every photon is phase-locked to the excitation laser with mutual-coherence time exceeding 3 seconds, which corresponds to a mutual-coherence length of one million kilometres. These results confirm the prediction for an ideal two-level system that each photon inherits fully the coherence properties of the excitation laser in this alternative photon generation process. The absence of linewidth broadening beyond our detection limits further verifies that spectral diffusion occurs on timescales of seconds or slower in our experiment which will permit feedback to stabilise the transition, e. g. using electric fields[24].

**Phase-shaping of single photon waveforms.** Methods currently available to control single photon wavepackets range from direct spectral filtering[25] to intra-photon phase encoding via photon transmission through electro-optic elements[26, 27, 28]. Wavepacket control during the photon generation process has only been achieved for trapped atoms inside optical cavities using multi-pulse sequences[29, 30]. The coherent nature of elastic scattering provides a means for coherent synthesis of single photon waveforms directly in the photon generation process without the need for spectral filtering or optical cavities. In order to demonstrate this ability, we use an electro-optic modulator (EOM) driven by a variable amplitude 200-MHz radio-frequency source (see Supplementary Methods) to encode the excitation laser field. Figure 2 presents two examples of coherent control of single photon waveforms by this technique. The first example of a synthesized laser waveform and the corresponding spectrum are displayed in Figures 2a and 2b, respectively. The temporal measurement in panel a is performed using a photodiode with 8-GHz bandwidth, while the spectrum in panel b is measured using a stabilized Fabry-Perot interferometer (see Methods). Figures 2c and 2d display another example for encoding the laser field with a more complicated pattern, where the component at the carrier frequency (zero detuning) is strongly suppressed. The single photon QD emission spectra for the two examples are presented in Figures 2e and 2f. The 200-MHz spectral spacing and the relative strength of the frequency components are dictated by the laser spectrum for each example, in addition to being weighted by the QD transition linewidth. The radiative linewidth for a measured excited state lifetime of 0.65 ns is indicated as gray shaded areas in panels e and f for comparison. In both cases, the strong antibunching observed in the intensity-correlation measurements is fully sustained (see Supplementary Note 2 and Supplementary Fig. S2). We note that while these examples are based on amplitude modulation of the laser field, encoding phase to each spectral component is possible in this scheme. These measurements demonstrate that, as a consequence of their mutual coherence with the laser field, the single photon waveforms can be synthesized deterministically without post-generation filtering which can lead to photon loss and degradation of the single photon quality.

**Indistinguishability of pulsed coherent single photons.** A key requirement of a photonic link in a quantum network is time-synchronized indistinguishable single photon streams. However, the uncorrelated environment fluctuations experienced by separate QDs under nonresonant excitation lead to variations in temporal and spectral overlap of photon wavepackets[31, 32]. Imparting mutual coherence on independent photons via a common excitation laser can therefore be a crucial advantage in achieving indistinguishable wavepackets. Therefore, we perform Hong-Ou-Mandel (HOM) style two-photon interference (TPI) experiments[33] of two QD photons scattered from the same QD at different times to quantify the indistinguishability[34] of synthesized photon waveforms with complex spectra. A schematic of the experimental setup is shown in Fig. 3a. Figure 3b displays the laser pulse train (top panel) used to generate coherent QD photons in the Heitler regime (bottom). The laser pulses are derived from a continuous wave laser by modulating the EOM transmission with independent control on the duration and the repetition rate sustained (see Supplementary Fig. S3). Here, the laser pulses are 500-ps long with a repetition rate of 300 MHz. For these settings the temporal response of the QD transition is visible as an exponential tail in the coherent conversion of laser pulses into single photons. Figure 3c shows the intensity-correlation measurement, i.e. $g^{(2)}(\tau)$, for these photons. The missing peak at zero time delay in coincidence detection evidences strong antibunching. For pure photonic states, a HOM-style TPI measurement in the vicinity of zero time delay is expected to follow

$$g_{\mathrm{HOM}}^{(2)}(\tau) = \left(T_1^2 + R_1^2\right)g^{(2)}(\tau) + 2R_1T_1\left(1 - \eta + \eta g^{(2)}(\tau)\right), \tag{1}$$

where $T_1$ and $R_1$ are the transmission and reflection coefficients of the first beamsplitter used to form the two input arms of the second beam splitter (cf. Fig. 3a). The variable $\eta$ quantifies the photon indistinguishability including any imperfections of the measurement apparatus. For fully indistinguishable ($\eta = 1$) photons $g_{\mathrm{HOM}}^{(2)}(\tau)$ reduces to $g^{(2)}(\tau)$, while for fully distinguishable ($\eta = 0$) photons Eq. (1) takes the form $g_{\mathrm{HOM}}^{(2)}(\tau) = \left(T_1^2 + R_1^2\right)g^{(2)}(\tau) + 2R_1T_1$. We use the polarisation of the input photons as means to vary $\eta$, such that parallel (orthogonal) polarisation leads to a maximum (minimum) value for $\eta$. These measurements, performed in conditions identical to those of Fig. 3c, are shown in Figs. 3d and 3e for orthogonal and parallel input polarisation states, respectively. In order to quantify the TPI fidelity we compare the data directly with the ideal case of $g^{(2)}_{HOM}(\tau) = g^{(2)}(\tau)$: the bottom panels of Fig. 3d and e show $\left(g^{(2)}_{HOM}(\tau)/g^{(2)}(\tau) - 1\right)$ for their respective HOM measurement. Integrating this normalised difference curve over one repetition period (~3.3 ns) centred around zero time delay, we can extract a raw contrast of the TPI experiment of $C_{HOM} = 1 - (A_{pa}/A_{or}) = 0.926 \pm 0.016$. Taking into account deviations from ideal experimental conditions, e.g. imperfect polarisation control and unbalanced interferometer arms, in the conventional way[34] we extract a corrected contrast of unity within the experimental uncertainty ($C_{cor} = 1.03 \pm 0.05$). However, the correction for the unbalanced interferometer is complicated by the coherence of the photons beyond the interferometer delay sustained (see Supplementary Notes 3 and 4 and Supplementary Fig. S4). Therefore, limiting the corrections to only those arising from imperfect polarisation control we report a lower bound on photon wavepacket indistinguishability of $0.96 \pm 0.04$. This value surpasses all reports on spontaneously emitted QD photons to-date and confirms the expected advantage of the coherently scattered photons replicating the laser coherence. A TPI measurement for higher excitation power is presented in Supplementary Fig. S5. We stress that the lack of correlation between the environment dynamics of multiple QDs will inevitably lead to the degradation of photon indistinguishability for the incoherently generated photons as is evident in all attempts to date[31, 32], while the coherent generation process reported here ensures the immunity of our time-synchronised coherent photon wavepackets to such effects.

**Discussion**

We have demonstrated the generation of fundamentally indistinguishable, coherent single photons from a solid-state quantum emitter with a high degree of control and flexibility in waveform synthesis. These phase-locked photons are ideally suited for quantum interference applications, which form the basis of quantum communication[18], linear optics quantum computation[35], and distant entanglement schemes[10, 11]. The price to pay for this unprecedented photon quality is excitation efficiency. To obtain a coherent fraction of about 0.9 in pulsed excitation, photons are probabilistically generated with 5-10% efficiency (1.5-3x10$^7$ photons/s for the HOM measurements presented here). While generation rates are lower than those achievable in deterministic excitation schemes, the trade-off between photon rates and photon quality is well justified for two main reasons: first, quantum interference applications can work well with probabilistic gates[11, 36], while the low photon quality reported so far[31, 32] rules out the use of QD photons in quantum information applications. Further, non-unity photon collection and detector efficiencies evidently render any measurement-based schemes probabilistic. Second, an entanglement rate scaling with the two-photon detection efficiency[11] is replaced advantageously by one-photon detection[10, 18], utilising the first-order coherence of the coherently scattered photons. The technique presented in the work is not limited to QDs and can be extended to other quantum systems in the optical domain as well as to superconducting circuits in the microwave domain. Spin-selective transitions of QDs will be utilized next in order to transfer the quantum mechanical state of a spin to frequency and polarisation degrees of freedom of a photon. The ability to tailor single photon waveforms is also advantageous in efficiently storing information in quantum memories with restricted spectral bandwidth. Building on these results we aim to realize a distributed network of multiple QD spins interfaced with waveform-synthesized and time-synchronized single photons – all phase-locked to a master laser.

## Methods

**Quantum dot sample structure.** We use a sample grown by molecular beam epitaxy containing a single layer of self-assembled InAs/GaAs QDs in a GaAs matrix and embedded in a Schottky diode for charge state control. The Schottky diode is formed by gating an n+-doped layer 40 nm below the QDs with a 5-6 nm thick partially transparent titanium layer evaporated on top of the sample surface. This device structure allows for deterministic charging of the QDs and shifting of the QD exciton energy levels via the DC stark effect. 20 pairs of GaAs/AlGaAs layers forming a distributed Bragg reflector extend to 205 nm below the QD layer for increased collection efficiency in the spectral region between 960 nm and 980 nm. Spatial resolution and collection efficiency are enhanced by a zirconia solid immersion lens in Weierstrass geometry positioned on the top surface of the device. The measurements reported here are repeated on two separate QDs for further confirmation. For both QDs the neutral exciton transition (one transition of the exchange doublet) was optically addressed. The wavelength of this transition was at 951 nm for both QDs.

**Resonance fluorescence measurement technique.** Resonant optical excitation and collection of the fluorescence from single QDs is achieved using a confocal microscope with 10:90 beam splitter head in a 4.2-K liquid helium bath cryostat. We excite and collect both

QD fluorescence and laser scattering along the same path via a 0.4 NA objective lens inside the cryostat. The QD transition dipole has a preferential in-plane orientation, independent of excitation orientation and polarisation, therefore we excite in a linear polarisation at ~45° to the transition polarisation and collect in the orthogonal polarisation. This ideally removes all laser at the cost of filtering half of the scattered QD photons. In practice, $10^7$ suppression of the laser reflection is achieved combining this cross-polarized detection with confocal rejection due to coupling to a single mode fibre. The contribution of the phonon sideband represents ~12% of the total QD emission in our sample. For TPI measurements the QD fluorescence is spectrally filtered via a 1600 gr/mm grating acting as a bandpass filter for the zero-phonon line. The experimental setup including the dark field detection scheme and the spectral filtering, together with an unfiltered emission spectrum is illustrated in Supplementary Figure S6. The laser background to signal ratio is well below 1% for all measurements presented. Under pulsed excitation the excitation power was chosen so that the elastically scattered photons exceed 90% of the total fluorescence (see Supplementary Note 3 and Supplementary Fig. S4). Working in this regime yields approximately one scattering event per 20 excitation pulses.

**Optical heterodyning of QD resonance fluorescence.** A continuous wave laser at optical frequency $\nu$ is split into two beams of different intensities where the weak laser beam excites the QD. The strong laser beam is shifted in frequency to $\nu + \delta\nu$ using acousto-optical modulation (AOM). In these experiments $\delta\nu$=210 kHz, achieved by frequency-shifting the laser twice using two AOM devices operating at 80.000 MHz and 79.790 MHz. This avoids any residual electronic pickup noise in the experiment in the spectral region of interest. The fields of the coherently scattered single photons $\vec{E}_S$ (signal) and the frequency-shifted laser $\vec{E}_{LO}$ (local oscillator) are combined via a beam splitter and detected by two balanced photodiodes (PD1,2 in Fig. 1). The total optical intensity $I_i$ on each PD contains a phase-invariant component comprising the sum of the two input intensities and a phase-dependent component oscillating at the difference frequency $\delta\nu$. $I_i$ has the mathematical form

$$I_i = \left|\vec{E}_S + \vec{E}_{LO}\right|^2 = \left|\vec{E}_S\right|^2 + \left|\vec{E}_{LO}\right|^2 + 2\vec{E}_S\vec{E}_{LO}\cos\left(\delta\nu \cdot t + \varphi_i(t)\right), \tag{2}$$

where $\phi_i(t)$ is the relative phase between the two optical fields on PD$_i$ at time $t$. Working with the difference of the PD signals suppresses the constant components in Eq. (2), while enhancing the beating signal amplitude. The PDs used for this measurement have 500 kHz detection bandwidth. The resultant power spectrum is governed by the dynamics of the relative phase, $\phi_i(t)$, in the course of the measurement. For a Gaussian lineshape the width of the power spectrum, $\Delta \nu_{FWHM}$, can be converted into a mutual-coherence time, $\tau_c$, via the relation $\tau_c = \sqrt{2\ln 2/\pi}/\Delta \nu_{FWHM} \approx 0.66/\Delta \nu_{FWHM}$. The mutual-coherence length is obtained using $c \times \tau_c \approx 0.66 \times c/\Delta \nu_{FWHM}$.

**Details of the Hong-Ou-Mandel interferometer.** To observe TPI we split the stream of QD photons into two arms on a nominally 50:50 beamsplitter (measured $T_1/R_1 \sim 1.3$). The delay between the two arms ($\Delta t_{delay} = 3.33$ ns) is adjusted so that two consecutive pulses overlap temporally at the second beam-splitter ($T_2 = 0.41 \pm 0.01$, $R_2 = 0.59 \pm 0.01$). The latter is fibre-based (single mode), which ensures spatial mode matching, but requires additional care for conserving the polarization of the input photons. We use custom-made polarisation controllers and half-wave plates to match the input polarizations. The half-wave plate shown in Fig. 3 allows us to perform TPI measurements for photons with orthogonal and parallel polarisations with the same setup. The degree of polarisation control in the interferometer is measured separately before each measurement for the two half-wave plate settings corresponding to interference of parallel and orthogonal polarisations, respectively, by sending through a narrowband single-mode laser. For the measurements presented in Fig. 3 we match polarisation in the interferometer to $p_{pa} = 0.97 \pm 0.02$ in the parallel case and measure a contrast of $p_{or} = 0.05 \pm 0.02$ in the orthogonal case.

**Instrument properties.** The laser-stabilized Fabry-Perot cavity has a resolution of 20 MHz and the two-photon detection system used for intensity-correlation as well as Hong-Ou-Mandel interference measurements has a temporal resolution of 600 ps. The lasers are stabilized to <2 MHz linewidth over few hours. Uncertainties stated for the interference contrast represent one standard deviation and are based on propagation of uncorrelated random errors.

## Acknowledgements

This work was supported by grants and funds from the University of Cambridge, EPSRC Science and Innovation Awards, the QIPIRC and EPSRC grant number EP/G000883/1. M.G. acknowledges the German Research Foundation (DFG) for financial support (grant number GE 2141/2-1) and Z. L. acknowledges the support of NSFC (grant number 60907027). We thank A. N. Vamivakas, J. M. Taylor, C.-Y. Lu, R. T. Phillips and A. J. Ferguson for fruitful discussions, as well as technical assistance.

## Contributions

C. M., M. G., C.H.H.S., C.L.G., J.H., Z.L. and M.A. contributed to conception, instrumentation, measurements and analysis. M.H. and E.C. grew the sample. All authors discussed the results and contributed to writing the manuscript.

## Competing financial interests

The authors declare no competing financial interests.

## Corresponding author

Correspondence to: Mete Atature (ma424@cam.ac.uk)

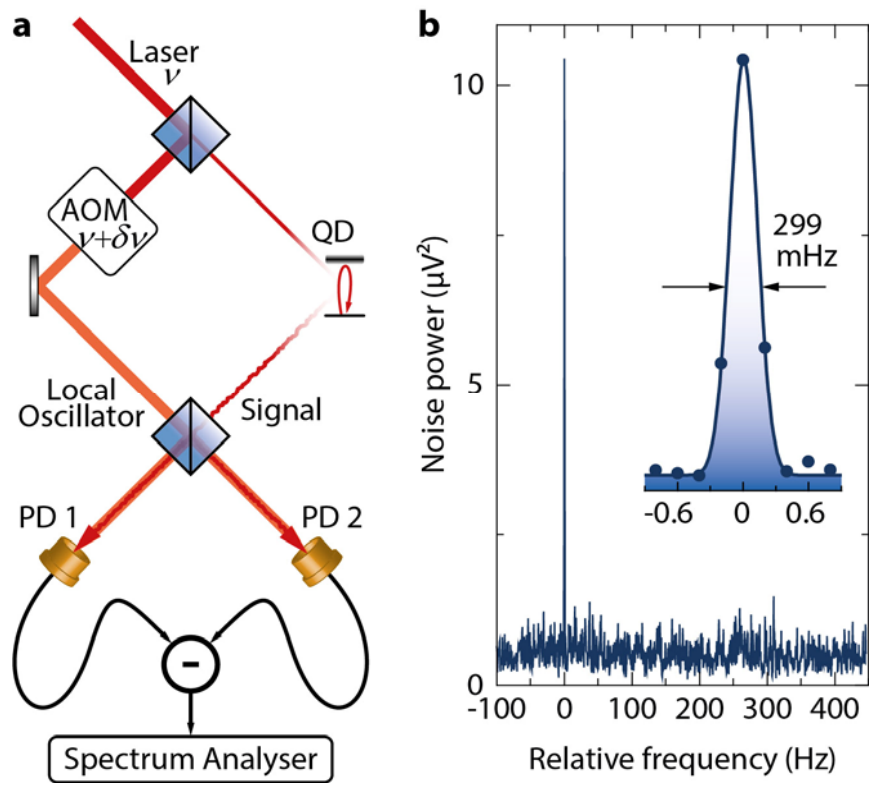

**Figure 1| Measurement of mutual coherence using optical heterodyning. a,** An illustration of the experimental arrangement used for optical heterodyning of QD resonance fluorescence. Acousto-optical modulation (AOM) provides the shift of the local oscillator frequency by $\delta\nu \approx$ 210 kHz. The outputs of the two photodiodes (PD1,2) are subtracted electrically and sent to a spectrum analyser. **b,** Typical power spectrum of the beating signal as a function of frequency relative to $\delta\nu$. Inset: high resolution spectrum for 5 seconds of continuous data acquisition. Spectra with sub-hertz resolution reveal phase coherence between QD photons and excitation laser on a time scale of seconds.

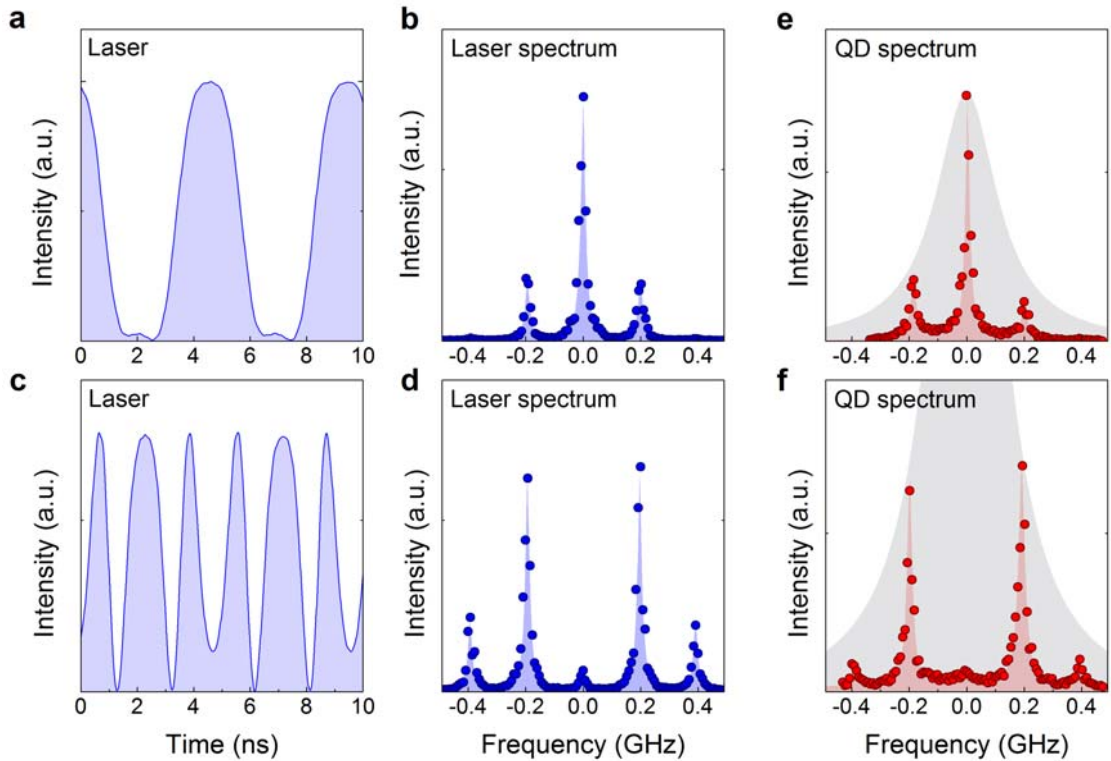

**Figure 2| Imprinting arbitrary waveforms on single photons. a,** The intensity of the excitation laser is modulated in time using an EOM driven with a 200 MHz sine wave. **b,** The spectrum of the modulated laser, measured through a Fabry-Perot cavity (resolution~20 MHz), shows sidebands at the modulation frequency. **c,d,** A different waveform modulation with the corresponding spectrum for the laser. This waveform is designed to suppress the spectral component at the original carrier frequency. **e,f,** Measured spectra of the QD photons generated from the two synthesized laser waveforms given in panels **a** and **c**, respectively. The elastically scattered photons replicate the laser spectrum within the linewidth of the transition which is indicated by the shaded grey area.

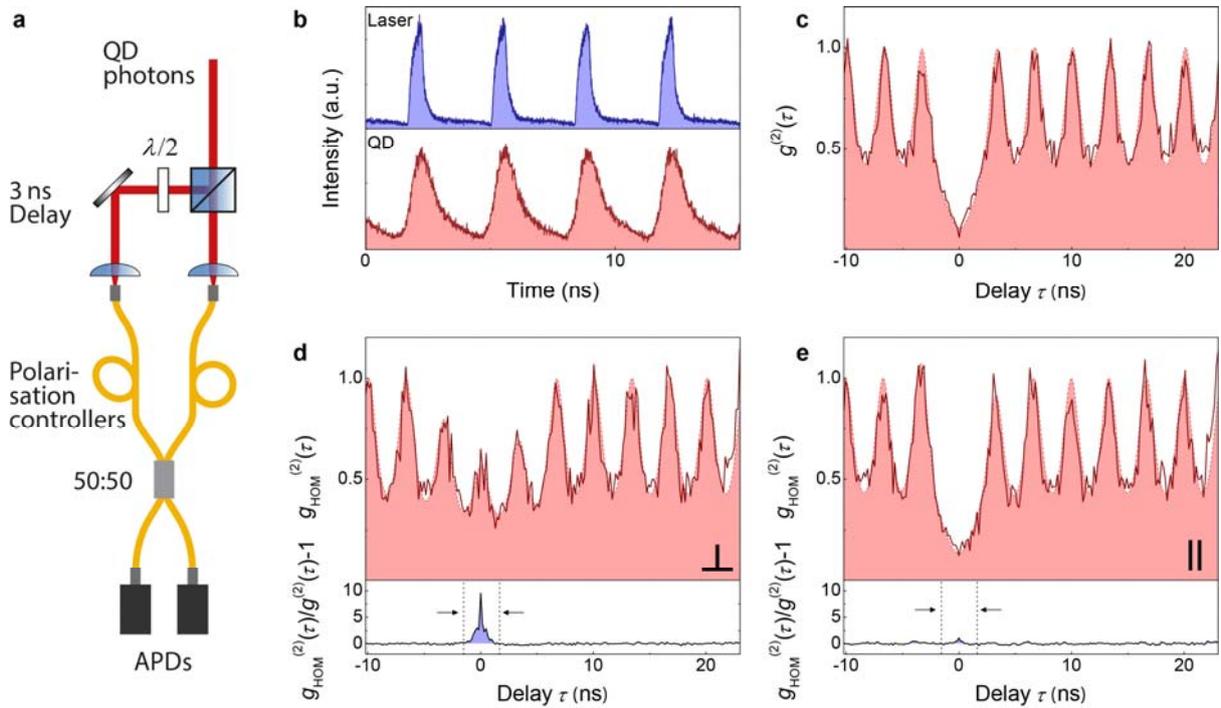

**Figure 3| Indistinguishability of waveform-synthesized photons. a**, Illustration of the experimental setup for two-photon interference from a single QD. **b,** Time-synchronized single QD photons (bottom panel) are generated by exciting the QD transition with laser pulses of 500-ps width and 300-MHz repetition rate (top panel). **c,** The intensity autocorrelation function for QD photons is recorded in a Hanbury-Brown and Twiss setup. A two-photon probability per pulse below 5% demonstrates the quantum nature of the emitter. **d, e** Two-photon interference (TPI) measurements for QD photons of orthogonal (**d**) and parallel (**e**) polarisation. The bottom panels display the normalized differences to the intensity autocorrelation in **c**, respectively. The lack of TPI for orthogonally polarized photons in **d** gives rise to coincidences at zero time delay, yielding a peak in the bottom panel. For photons of parallel polarisation, the absence of coincidences around zero time delay reflects successful TPI in **e**. The ratio of the areas in the bottom panels yields a raw visibility of 0.926. Experimental data in **c**, **d**, **e** are shown as continuous dark red curves, while a simulation for each case based on system parameters is shaded in light red. The data is recorded in 162-ps time bins with 600-ps timing resolution.